\documentclass[a4paper,9pt]{article}

\usepackage[utf8]{inputenc}
\usepackage[T2A]{fontenc}
\usepackage{epsfig}
\usepackage{indentfirst} 
\usepackage{setspace} 
\usepackage{multicol} 
\usepackage{amsmath} 
\usepackage{textcomp} 
\usepackage{amssymb} 
\usepackage{upgreek} 
\usepackage{tikz} 
\usepackage{pgfplots} 
\usepackage{capt-of}

\usepackage{float}

\pgfplotsset{width=14cm,compat=1.3}

\usepackage{helvet}
\usepackage{fancyhdr}


\usepackage{geometry}
\usepackage{gnuplottex}

\usepackage{titlesec}

\titleformat{\chapter}[display]
{\filcenter}
{\thechapter}
{20pt}
{\bfseries}{}

\titleformat{\section}
{\large\bfseries}
{\thesection .}
{0.5em}{}

\titleformat{\subsection}
{\normalsize\bfseries}
{\thesubsection .}
{0.5em}{}
\titlespacing*{\chapter}{0pt}{-30pt}{20pt}
\titlespacing*{\section}{0pt}{1ex plus 0.5ex minus 0.2ex}{1ex plus 0.5ex minus 0.2ex}
\titlespacing*{\subsection}{0pt}{1ex plus 0.5ex minus 0.2ex}{1ex plus 0.5ex minus 0.2ex}

\usepackage{abstract}

\begin{document}
	\twocolumn[	
	\center{\textbf{Some thermodynamic features in space-times with violation of causality}}
	
	\center{A.A.Nikitenko}	
	\center{Bogoliubov Laboratory of Theoretical Physics, JINR,} 
	\center{Dubna, Moscow region, Russia}
		
	\begin{abstract}
	General relativity does not prohibit the existence of space-times that describe time travel. Consideration of such spaces gives rise to a lot of questions and paradoxes, among which there are thermodynamic ones. This paper considers two situations that describe time travel, and explains why their existence does not mean that time machines are prohibited.
	\end{abstract}
	]
	In this article, we will discuss some of the properties of time machines allowed by general relativity, related to the second law of thermodynamics. The thermodynamic behavior of travelers depends on the geometric properties of their world lines. To consider this dependence, we introduce the relevant definitions.
	
	The space-time is called a time machine \cite{Krasnikov:1} if it containes a point $p$ such that $J^{-}(p) \bigcap J^{+}(p)\neq p$, where $J^{-}(p)$ is the causal past, and $J^{+}(p)$ is the causal future of the point $p$. 
	
	
	
	
	
	Consider two different types of time travel.
	
	--- A traveler at some moment of his proper of time in a certain initial state, after some time, according to his own clock, completely (precisely) restores his original state. Following \cite{Rovelli:2} we will geometrically describe such a situation by a closed non-spacelike curve. Mathematically, such a curve is described by a map $ \gamma(\lambda)\text {:} (a; b) \longrightarrow M $ such that $\gamma(\lambda) = \gamma(\lambda + l)$ for an arbitrary value of $\lambda$, where $l$ is some constant. That is, the world line of the traveler is a closed loop.
	
	--- As the second situation, consider the one in which the time traveler returns to the past, to a small neighborhood of the starting point and has the opportunity to send a signal from it to his younger version. To describe such situations, it is appropriate, without loss of generality, to resort to an idealization. Namely, instead of considering a piecewise non-spacelike curve: $\gamma_{1}(\tau) \cup \gamma_{2}(\lambda)$, consisting of the traveler's world line $\gamma_{1}(\tau) $ and the world line of the signal $\gamma_{2}(\lambda)$, which is sent by the old version of the traveler to its younger version, we will consider one curve --- the world line of the traveler $\gamma(\lambda)$, neglecting a small segment containing the signal world line. In our idealization the neighborhood in discussion will correspond to the point at which the traveler's world line experiences self-intersection. We emphasize that in this a situation, in the neighborhood in discussion, there are two different world lines - the world line of the traveler and that of the signal sent by the old traveler to his younger version. For a geometric description of such a situation, we will use the term: self-intersecting non-spacelike curve $\gamma(\lambda) \text {:} (a; b) \longrightarrow M $, with the self-intersection condition $\gamma(\lambda_{1}) = \gamma(\lambda_{2}) $, implying that the curve physically describes this particular idealization. It may still not seem quite obvious how these two situations differ. Therefore, we will explain that in more detail.
		
	Closed non-spacelike curves differ from non-spacelike curves with self-intersection. The difference is that the system described by such a world line as a result of its evolution necessarily returns to its original state. For expample consider the following situation. Let a glass beaker have a closed world line and first stands on a table. Then it falls off the table to the floor and breaks. The closedness of its world line means that after some time the particles of which it consisted will be collected again in the glass and will again be on the table in their initial state. In such a situation, the traveler never leaves the non-causal region.	
					
	In \cite{Rovelli:2} Rovelli consideres closed timelike curves as a special case of time machine. He introduces the proper time along the trajectory of the traveler, and  ascribes the proper entropy to him, which he considers as a function of the proper time $S(\tau)$. Then, after the traveler makes a complete revolution around the loop, Rovelli equates the entropy values of the old traveler and his younger version, referring to the fact that in the loop, proper time and entropy must be periodic.
	
	\textit{Note:} Note that time travel described by closed timelike curves was considered both in fiction \cite{Harrison:3} and in general relativity \cite{Novikov:4}, and among them there are indeed ones, in which the second law of thermodynamics is not satisfied. In the work \cite{Novikov:4}, the authors note that in the case of time machines with closed non-spacelike curves, a traveler moving along such a loop can interact with bodies that have come into the non-causal region from the outside. As a result of this interaction, it can decrease its entropy; therefore, along the loop, its entropy may not be a monotonic function of its proper time. That is, even among time machines of this type, there are also those for which the second law of thermodynamics is fulfilled. In the sense that the entropy of such a traveler can be periodic along the loop without contradicting the second law of thermodynamics.

	The flaw in Rovelli's analysis is that, considering a special case of a time machine described by a closed timelike observer curve, he draws a conclusion both for time travel described by closed timelike curves, and for those described by self-intersecting, but not closed, timelike curves. Ref. \cite{Rovelli:2} states "Since no function can uniformly increase around a circle, we can never "travel to the past"\, in the sense of arriving to the past having memory of the future, having counted the oscillations of our clock, being in a position of acting different that what we did, or similar."\, From which Rovelli concludes:
	"The closed timelike curves of general relativity generate no paradoxes. But they do not allow us to travel to past in the thermodynamical sense | for instance being in the past having memory of the future."\, The above quote from Rovelli is valid for time travel, which is described by some kinds of space-times with closed timelike curves. Below we show that for those time trips that are described by space-times with self-intersecting timelike curves, this does not hold, since the entropy along such a traveler's world line can increase monotonically.	
	
	To show that Rovelli's analysis \cite{Rovelli:2} will not work consider now a situation described by self-intersecting non-spacelike curves. Indeed: despite the idealization described above, the violation of causality in such a situation occurs due to the transmission of a signal from the old version of the traveler to his younger version. In this case, the traveler himself does not return to his previous state. Therefore, the entropy of a traveler along a self-intersecting non-spacelike curve can increase monotonically without violating the second law of thermodynamics and additional assumptions: if $\tau_{2}$ and $\tau_{1}$ are two times of the traveler's proper time such that $\tau_{2}> \tau_{1} $, then always $S(\tau_{2}) \geqslant S(\tau_{1}) $, along the whole curve. In other words, there are two different people in the vicinity of the self-intersection point: the old traveler and his younger version. In the situation described, the older traveler does not transform into his younger version. Its entropy will always be greater than the entropy of its young version. It is the possibility of such self-intersecting causal curves that is overlooked in \cite{Rovelli:2}.

	
			
	In the monograph \cite{Krasnikov:1}, Chapter 6, Figure 5, the twisted Deutsch-Politzer spacetime is used to discuss the machine builder's paradox. In order to demonstrate the behavior of the entropy of a traveler in a space-time containing self-intersecting timelike curves, this space is also quite suitable. We use it to show that Rovelli's analysis fails in the case of self-intersecting timelike curves. Instead of the isotropic curve shown in the example in Figure 6.5 of the monograph, we will consider the timelike curve $\gamma(\tau)$.
					
	\begin{figure}[H]
	\centering
	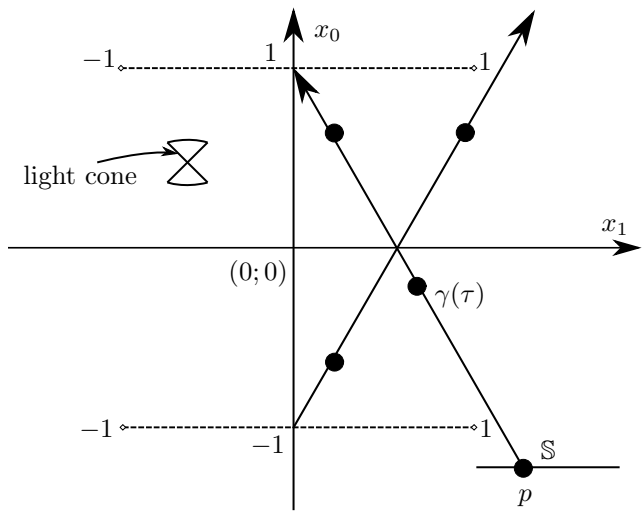
	\caption{The twisted space-time of Deutsch-Politzer. The traveler, who is initially on the surface $\mathbb{S}$, flies into the non-causal region. The circles indicate travelers with clocks at certain points in time. The light cones are directed at an angle of $ 45^{\circ}$. The other two spatial coordinates $x_{2}$, $x_{3}$ are omitted in the diagram.}
	\label{Time_Machine}
	\end{figure} 	

	To simplify the consideration, we will carry out all the reasoning for the (1 + 1) -dimensional Deutsch-Politzer space, presenting the situation (3 + 1). This simplifies the technical side of the matter and does not lead to loss of generality. The twisted Deutsch-Politzer space is shown in Fig. \ref{Time_Machine}.	It is constructed as follows: two intervals $-1 \leq x_{1} \leq 1 $ are removed for $x_{0}=1$ and $-1\leq x_{1} \leq 1 $ for $x_{0}=-1$. The upper bank of the upper cut is identified with the lower bank of the lower cut after one of them is rotated by $180^{\circ} $, similarly the lower bank of the upper cut is identified with the upper bank of the lower cut after one of them is rotated by $180^{\circ}$.
	
	Consider a traveler moving along the timelike curve $\gamma(\tau) $ in Fig. \ref{Time_Machine}. Obviously, some $\gamma(\tau)$ violate the causality condition and thus describe time travel. To pose thermodynamic questions about entropy, we must first of all agree on the values of the entropy of which objects we are going to compare. Since Rovelli is interested in the thermodynamic properties of the clock of a traveler moving along a timelike curve, we will compare the entropy of his clock at different times of $\tau_{1} <\tau_{2}$ proper time.	

	First, the traveler is in the globally hyperbolic region of space-time on the surface $\mathbb{S}$. Then he enters the non-causal area, moves in it to the lower edge of the upper cut. It comes out of the upper edge of the lower cut and has the ability to interact with itself (self-intersection). At the self-intersection point of the $\gamma(\tau)$ curve, the old traveler can interact with his younger version. The entropy of the old traveler, as it should be, is greater than the entropy of the young one. Therefore, timelike curves of this type, violating causality, do not contradict the second law of thermodynamics.
	
	\section*{Acknowledgements}
	I am grateful to S.V. Krasnikov and A.B. Arbuzov for helpful discussions.

\end{document}